\RequirePackage[2020-02-02]{latexrelease}
\documentclass[aps,prl,twocolumn,floats,10pt]{revtex4}
\usepackage{graphicx}
\usepackage{amsmath}
\usepackage{float}

\begin{document}

\title{Causal Contradiction is absent in Antitelephone}

\author{T. Zh. Esirkepov}
\affiliation{Kansai Photon Science Institute, National Institutes for Quantum and Radiological Science and Technology (QST), 8-1-7 Umemidai, Kizugawa, Kyoto 619-0215, Japan}

\begin{abstract}
Thought experiments in the ``antitelephone'' concept
with superluminal communication
do not have causal contradiction.
\end{abstract}

\maketitle


Many physical phenomena exhibit apparent superluminal motion.
A light spot is an everyday example: it can move with an arbitrary speed.
However, each moment of time it consists of different photons.
Thus, it does not transfer matter, energy and momentum
from one place of its appearance to another.
Consequently, the events of its appearance in different places
are not causally dependent, despite the fact that somebody
or something controls the flow of photons by casting
the light spot in different directions.
In the same manner occur superluminal modulations in various physical media.
For example, the phase velocity of electromagnetic waves in plasma
can be superluminal.

There is a vast discussion on the possibility of superluminal communications in the literature,
e. g., \cite{SciPop1}.
In popular science there is a belief that hypothetical superluminal communications lead to causal contradiction. Supposedly it allows a situation, where a message is sent to the past and modifies the sender intention. One position is that this causal contradiction alone makes superluminal communications impossible. Another position more subtly states that superluminal communications do not transfer information, thus sending it to the past cannot modify already happened events.

Causal contradiction due to superluminal communication is often illustrated with thought experiments in the framework of special relativity theory. One famous example is so called ``tachyonic antitelephone'' \cite{SciPop2}.
Here we show that the antitelephone setting does not
have causal contradiction.

We avoid using the term ``signal'' for superluminal communication
reserving it only for 
messages travelling not faster than $c$, the speed of light in vacuum.
We do not discuss the possibility or nature of superluminal communications.
The mathematical model described below
does not generate arguments in favour of their feasibility.

We assume dimensionless values, where $c=1$.
More precisely, one can assume that the unit of time is $c\cdot$second,
then velocity becomes dimensionless (normalized to $c$).
We consider only inertial reference frames.

\begin{figure}
\includegraphics[width=0.8\columnwidth]{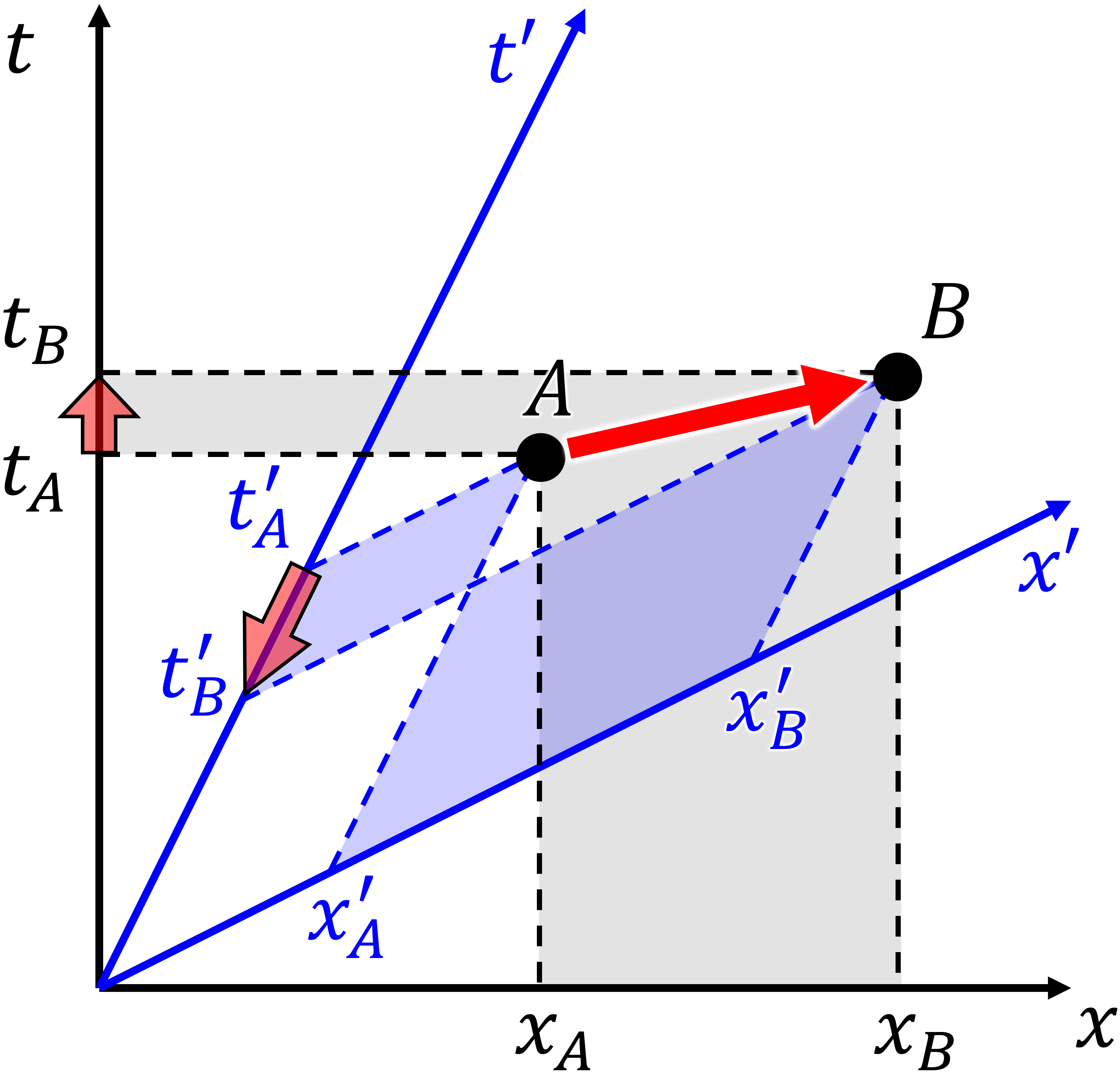}
\caption{
Minkowski diagram of the superluminal communication
from $\{t_A,x_A\}$ to $\{t_B,x_B\}$.
Semi-transparent outlined arrows show a projection of the world line of the
superluminal communication onto the temporal axes $t$ and $t'$.
World lines of light (not shown) are lines at $\pm 45^{o}$
with respect to the spatial or temporal axis. Hatched areas are for convenience.
\label{fig-1}
}
\end{figure}

In the simplest thought experiment, a superluminal communication with the speed
$V > 1$ is sent from the point $x_A$ at the moment of time $t_A$
and then reaches the point $x_B > x_A$ on the $x$-axis at the moment of time $t_B$.
In the 2-dimensional Minkowski space,
these two events are represented by two 2-vectors:
$e_A = \{ t_A, x_A \}$ and $e_B = \{ t_B, x_B \}$.
The time delay between these two events is $\Delta t = t_B - t_A = (x_B - x_A)/V > 0 $.
In the reference frame, moving with the velocity $\beta$, $|\beta|<1$,
in the direction of the $x$-axis,
the coordinates of these events can be deduced from
the Minkowski diagram in Fig. \ref{fig-1}.
It illustrates a Lorentz transformation of the 2-dimensional Minkowski space
$(t,x)\rightarrow (t',x')$,
which is a linear transfromation represented by the matrix
\begin{equation}\label{eq-M2}
{\cal M} = \left[
\begin{array}{cc}
\gamma & -\beta\gamma \\
-\beta\gamma & \gamma
\end{array}
\right] ,
\end{equation}
where $\gamma=(1-\beta^2)^{-1/2}$.
We note that all derivations below are simple linear algebra,
completely equivalent to the calculation of lengths of various
segments and their projections onto $t$- and $t'$-axes
in the Minkowski diagram.

In the moving reference frame, the coordinates of the event $e=\{t,x\}$ are
\begin{equation}\label{eq-tAM}
\{t',x'\} = {\cal M}\!\cdot\!\{t,x\} = \{\gamma(t-\beta x),\gamma(x-\beta t)\} .
\end{equation}
The time period between the events $e'_A$ and $e'_B$
is the first component of the 2-vector
$e'_B-e'_A = {\cal M}\!\cdot\!( e_B-e_A )$:
\begin{equation}\label{eq-DtM}
\Delta t' = t'_B-t'_A = \gamma (1-\beta V) \Delta t .
\end{equation}
If  $V>1⁄\beta$, the time delay is negative: $\Delta t' < 0$,
as shown in Fig. \ref{fig-1}.

The described thought experiment shows that, according to clocks in the moving reference,
one can receive a superluminal communication before its emission.
Sometimes this is regarded as a sign of causal contradiction.
The main subtle aspect of such a conclusion is the implicit assumption that the clocks in the each of the reference frames are synchronized using signals with the speed of light.
This synchronization is ``encoded'' in the metric of the Minkowski space.
It is not surprising that, if we synchronize clocks with a ``slow'' signals, the recorded timing for faster communications will be seen as ``incorrect''.
In order to have a causal contradiction, it is necessary to demonstrate a loop,
where superluminal communication can remove its cause.

\begin{figure}
\includegraphics[width=0.8\columnwidth]{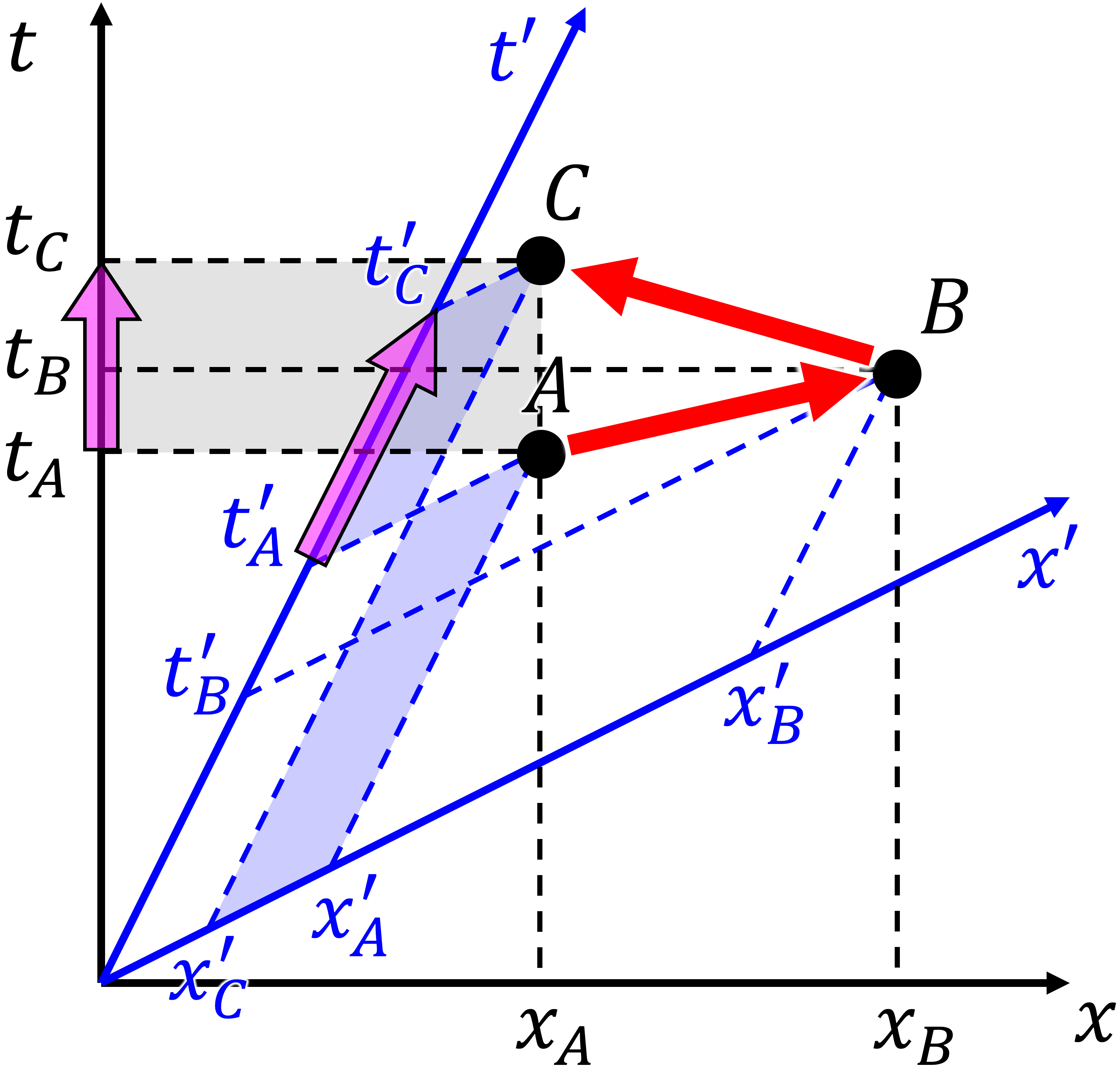}
\caption{
Minkowski diagram of the loop of superluminal communications
from $\{t_A,x_A\}$ to $\{t_B,x_B\}$, then to $\{t_C,x_C\}$, where $x_C=x_A$.
Semi-transparent outlined arrows show a projection of 
the vector $\overset{\rightarrow}{AC}$ (the result of superluminal communication loop)
onto the temporal axes $t$ and $t'$.
Hatched areas are for convenience.
\label{fig-2}
}
\end{figure}

We consider an extension of the above described thought experiment, 
where the superluminal communication is looped, Fig. \ref{fig-2}.
Two emitters, $A$ and $B$, are placed at the points $x_A$ and $x_B$, respectively,
along the $x$-axis.
We assume that when an emitter receives a superluminal communication, 
it sends a superluminal communication to another emitter with a negligible delay.
The speed of communication from $A$ to $B$ is $V_A$ and
from $B$ to $A$ is $V_B$.
We consider the following sequence of events.
Event $e_A = \{ t_A, x_A\}$: the emitter $A$ sends a communication to $B$.
Event $e_B = \{ t_B, x_B\}$: the emitter $B$ receives the communication from $A$.
The time period between these events is
\begin{equation}\label{eq-tBA}
t_B-t_A = \frac{x_B-x_A}{V_A} .
\end{equation}
Event $e_C = \{ t_C, x_C\}$, $x_C=x_A$:
the emitter $A$ receives in reply the communication from $B$.
The time periods between the events $B$ and $C$ is
\begin{equation}\label{eq-tCB}
t_C-t_B = \frac{x_B-x_A}{V_B} ,
\end{equation}
and between the events $A$ and $C$ is
\begin{equation}\label{eq-tCA}
t_C-t_A = (x_B-x_A)\frac{V_A+V_B}{V_A V_B}.
\end{equation}
The sign in the right hand side of Eq. (\ref{eq-tCB})
indicates that the reply is send in the opposite direction
with respect to the $x$-axis
(here we assume that both quantities $V_A$ and $V_B$ are positive).
According to Fig. \ref{fig-2}, $x_B > x_A$,
thus $t_C > t_B > t_A$.

We consider these events in the moving reference frame,
which moves with the velocity $\beta$, $|\beta|<1$, along the $x$-axis, Fig. \ref{fig-2}.
In this frame, the
time period between the events $e_A$ and $e_B$ 
is given by the first component of the 2-vector $e'_B-e'_A = {\cal M}\!\cdot\!(e_B-e_A)$:
\begin{equation}\label{eq-DtAM}
t'_B-t'_A = \gamma(x_B-x_A)\frac{1-\beta V_A}{V_A},
\end{equation}
which is the same as Eq. (\ref{eq-DtM}) if $V$ is substituted by $V_A$.
Calculating the coordinates of the events $e_B$ and $e_C$ in the moving reference frame,
we obtain the corresponding time periods between the events:
\begin{align}
&
t'_C-t'_B = \gamma(x_B-x_A)\frac{1+\beta V_B}{V_B},
\label{eq-tCBM}
\\
&
t'_C-t'_A = \gamma(x_B-x_A)\frac{V_A+V_B}{V_A V_B} = \gamma(t_C-t_A).
\label{eq-tCAM}
\end{align}

We can disregard oddities in the timings of pairs of events (which are not loops), since, as noted above, the usage of the time quantity in the Lorentz transformation implies that 
all the clocks are synchronized via signals with the speed of light.
We see from Eqs. (\ref{eq-tCA}) and (\ref{eq-tCAM}) 
and also from Fig. \ref{fig-2}
that in both reference frames, 
the time period between the events $e_A$ and $e_C$ is positive,
so that the reply comes later than the inquiry. 

The conclusion that loops have positive time span is valid in the general case,
when one or both emitters move, irrespective of the motion direction of the 
moving reference frame.
Following popular science approach, we consider a thought experiment,
where Alice and Bob are on different spaceships flying in different directions
with different (not superluminal) velocity. They communicate with each other
generating the following sequence of events 
represented by the 4-vectors in the Minkowski space.
Alice sends a communication:
\begin{equation}\label{eq-eA}
e_A=\{t_A,x_{A1},x_{A2},x_{A3}\} .
\end{equation}
Bob receives the communication and immediately replies:
\begin{equation}\label{eq-eB}
e_B=e_A+t_{AB}\{1,V_{AB1},V_{AB2},V_{AB3}\} . 
\end{equation}
Here $V_{ABi}$, $i=1,2,3$, are the spatial components
of the velocity of the communication from Alice to Bob.
Alice receives the reply:
\begin{equation}\label{eq-eC}
e_C=e_B+t_{BC}\{1,V_{BC1},V_{BC2},V_{BC3}\} . 
\end{equation}
Here $V_{BCi}$, $i=1,2,3$, are the spatial components
of the velocity of the communication from Bob to Alice.

Irrespective of the motion of Alice and Bob,
we can assume that the velocity vector of the moving reference frame
$\vec\beta$ , $|\beta|<1$,
defines the direction of the $x$-axis.
Of course, the events of Eqs. (\ref{eq-eA})-(\ref{eq-eC})
must be (re)defined with respect to that axis.
The matrix representing the Lorentz transformation of
the 4-dimensional Minkowski space to the moving reference frame is
\begin{equation}\label{eq-M4}
{\cal M} = \left[
\begin{array}{cccc}
\gamma & -\beta\gamma & 0 & 0 \\
-\beta\gamma & \gamma & 0 & 0 \\
0 & 0 & 1 & 0 \\
0 & 0 & 0 & 1
\end{array}
\right] .
\end{equation}

We consider the difference between the events
$e_C$ and $e_A$ in the moving reference frame,
\begin{align}
&
e'_C-e'_A = {\cal M}\!\cdot\!( e_C-e_A ) =
\nonumber\\
&
\big\{
\gamma\left[ t_{AB} + t_{BC} - \beta\left( t_{AB}V_{AB1} + t_{BC}V_{BC1} \right) \right] ,
\big. 
\nonumber\\
&
\gamma\left[ t_{AB}V_{AB1} + t_{BC}V_{BC1} - \beta\left( t_{AB} + t_{BC} \right) \right] ,
\nonumber\\
&
\big.
t_{AB}V_{AB2} + t_{BC}V_{BC2}, \;\; t_{AB}V_{AB3} + t_{BC}V_{BC3}
\big\}
.
\label{eq-eCAM}
\end{align}
The first component of the resulting 4-vector of Eq. (\ref{eq-eCAM})
is the time between the events
$e_C$ and $e_A$ in the moving reference frame,
\begin{equation}\label{eq-tAC-imp}
t'_{AC} = \gamma\left[ t_{AB} + t_{BC} - \beta\left( t_{AB}V_{AB1} + t_{BC}V_{BC1} \right) \right] .
\end{equation}

In order to receive the reply at the point $e_C$,
Alice must somehow arrive to this point from $e_A$.
We assume that she moves with some (not superluminal) speed $V_{AC}$:
\begin{align}
e_C=e_A+t_{AC}\{1,V_{AC1},V_{AC2},V_{AC3}\} ,
\label{eq-eC2} \\
|V_{AC}| = \sqrt{V_{AC1}^2+V_{AC2}^2+V_{AC3}^2} \le 1 .
\label{eq-VAC}
\end{align}
The consistency of Eqs. (\ref{eq-eC}) and (\ref{eq-eC2}),
which ensures that Alice receives the reply,
implies a quadruple of relationships
\begin{align}
&
t_{AC} = t_{AB} + t_{BC}, \nonumber\\
&
t_{AC}V_{ACi} = t_{AB}V_{ABi} + t_{BC}V_{BCi}, \; i=1,2,3.
\label{eq-4rel}
\end{align}
Using these relationships, we obtain from Eq. (\ref{eq-tAC-imp}):
\begin{equation}\label{eq-tAC-exp}
t'_{AC} = \gamma (1-\beta V_{AC1}) t_{AC} .
\end{equation}
Thus, $t'_{AC}>0$ if $\beta < 1$, $t_{AC} > 0$, and $|V_{AC1}| \le 1$.

The condition that $t_{AC}$ is positive
can follow from the requirement that 
both $t_{AB}$ in Eq. (\ref{eq-eB}) and $t_{BC}$ in Eq. (\ref{eq-eC}) are positive.
In its turn, this requirement indicates that
in the initial (steady) reference frame,
the recipients are in the right place and time
and the corresponding communications are accomplished.
We note that it is easy to arrange an impossible communication
even with not superluminal messages, e. g.,
when the world lines of the recipient and message do not intersect.

The structure of Eq. (\ref{eq-tAC-exp})
is the same as of Eqs. (\ref{eq-DtM}) and (\ref{eq-DtAM}),
as if the sequence of superluminal communications
\begin{equation}\label{eq-seq-ABC}
e_A\;\;\underset{V_{AB}}{\longrightarrow}\;\; e_B\;\;\underset{V_{BC}}{\longrightarrow}\;\; e_C
\end{equation}
were replaced by the only not superluminal communication (signal)
\begin{equation}\label{eq-seq-AC}
e_A\;\;\underset{V_{AC}}{\longrightarrow}\;\; e_C
,
\end{equation}
which Alice ``carries'' with her from the point $e_A$ ($e'_A$) to $e_C$ ($e'_C$).
The same observation concerns Eq. (\ref{eq-tCAM}),
where $t_C-t_A$ (or $t'_C-t'_A$) is just the time
of changing the state of the immobile emitter $A$.


The conclusion can be formulated as a theorem:
if there is an inertial reference frame where 
Alice receives in reply the communication from Bob later
than she sends her own communication to him,
then this is true in any inertial reference frame,
irrespective of the velocity of the communications.


The author thanks Dr. A. Bierwage, Dr. J. K. Koga and Dr. A. S. Pirozhkov for discussion.



\begin{thebibliography}{9}

\bibitem{SciPop1}
 G. Feinberg, ``Possibility of faster-than-light particles''.
{\it Physical Review}, {\bf 159} (5), 1089 
(1967). 

\bibitem{SciPop2}
G. A. Benford, D. L. Book, and W. A. Newcomb,
``The Tachyonic Antitelephone''.
{\it Phys. Rev. D} {\bf 2}, 263 (1970).

\end{thebibliography}
\end{document}